\documentclass[useAMS,usenatbib]{mn2e}

\usepackage{graphicx,times}
\usepackage{color}

\def\aj{{AJ}}

\def\red#1{{{}#1}}
\def\blue#1{{#1}}
\def\r1{$r_1$}
\title[Stellar-mass black holes in M22]{MOCCA code for star cluster simulations -- III.  Stellar-mass black holes in the globular cluster M22}
\author[D.C. Heggie and M. Giersz]
{Douglas
  C. Heggie$^{1}$
  \thanks{E-mail:
 d.c.heggie@ed.ac.uk (DCH); mig@camk.edu.pl (MG)} and Mirek Giersz$^{2}$
\\
$^1$School of Mathematics and Maxwell Institute for Mathematical
Sciences, University of Edinburgh, King's Buildings, Edinburgh EH9
3JZ, UK\\
$^{2}$Nicolaus Copernicus Astronomical Centre, Polish Academy of Sciences, ul. Bartycka 18, 00-716 Warsaw, Poland\\
}
\begin{document}

\date{Accepted \ldots. Received \ldots; in original form \ldots}

\pagerange{\pageref{firstpage}--\pageref{lastpage}} \pubyear{2002}

\maketitle

\label{firstpage}

\begin{abstract}
Using a Monte Carlo code, we construct a dynamic evolutionary model of
the Galactic globular cluster  M22 (NGC6656).   The initial conditions
are chosen so that, after about 12Gyr of stellar and dynamical
evolution, the model is an approximate fit to the surface brightness
and velocity dispersion profiles of the cluster, to its mass function,
and to the current binary fraction.  Depending on the distribution of
black hole natal kicks, we predict that the present-day population of
stellar-mass black holes ranges from about 40 (no kicks) down to essentially
zero (kicks distributed like those of neutron stars). { Provided that
natal kicks do not eject all new black holes, it is suggested that
clusters with a \red{present-day} half-mass relaxation time above about 1Gyr are the
ones that may still retain a\red{n appreciable} population of black holes.}
\end{abstract}


\section{Introduction}


For a long time, discussion of the role of black holes in globular
clusters has been dominated by the theme of intermediate-mass black
holes.  While no-one would doubt that stellar-mass black holes once
existed in these objects, one reason for their neglect is a
long-standing theoretical prediction \citep{KHM1993,SH1993} that there
should be virtually none at the present day.  As we shall see, this
view is
in need of revision, but in the meantime interest in the population of
stellar-mass black holes in globular clusters was sustained by two
ideas.  The first is the realisation that they may be a prolific
source of black-hole binaries and thus of sources of gravitational
radiation
\citep{PZM2000,B2006,BSRB2006,MS2009,BBK2010,DBGS2011,Ta2013}.  The
second is the role that stellar-mass black holes play in the evolution
of cluster cores \citep{MPPZH2004,Hu2007,MWDG2008}.

This paper focuses on the black hole population itself, and, in
particular, how many are to be expected at the present day in one
particular old
globular cluster.  Thanks to software advances over many years, it is
now quite straightforward to perform simulations of star clusters and
to study the evolution of the black hole population directly.  What is
still hard, however, is to do this with models which resemble globular star
clusters.  The most sophisticated direct $N$-body techniques have been
applied to this problem \citep[and several previous references by
other authors]{Aa2012}, but the restriction to systems which initially
possessed only of order $10^5$ stars vitiates their direct application to all except
the least populous globular
clusters.  There are two solutions, one being to {\sl scale} the
results of $N$-body models, provided that this can be done in a way
which preserves the time-scales of the main evolutionary processes at
work; the paper by \citet{SH2013} is an example, very relevant to the
scientific aims of the present paper, and we return to it in our
discussion (Sec.\ref{sec:bh-and-evolution}).  The second solution is the use
of Monte Carlo codes, which are not restricted to small values of $N$,
though they are less free of assumptions and approximations, and require cross-validation with $N$-body results in
the range of $N$ where the two techniques overlap \citep{GHH2008,
GHHH2013}.  A Monte Carlo code is the main tool adopted in the present
paper, but an independent code has also been applied to a similar
problem by \citet{MUFR2013}.

\begin{figure}
{\includegraphics[height=12cm,angle=0,width=9cm]{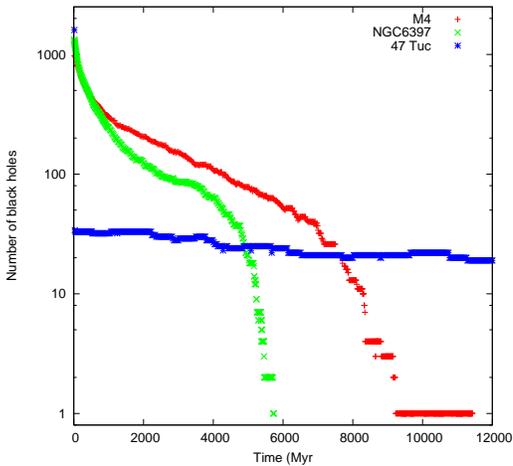}}
    \caption{Number of stellar-mass black holes in published Monte
    Carlo models of three globular clusters.  For the model of 47 Tuc,
    where natal kicks were applied, the results drop abruptly from the
    number initially created (about 1600).}
\label{fig:nbh}
  \end{figure}

Even with a Monte Carlo code, however, different sets of initial
conditions will give different answers for the number of stellar-mass
black holes expected to survive to the present day.  In previous papers
\citep{HG2008,GH2009,GH2011} Monte Carlo evolutionary models for three
clusters are described: M4, NGC6397 and 47 Tuc.  The number of black
holes throughout the evolution, which was not discussed much in these papers, is  plotted in Fig.\ref{fig:nbh}.  In
our model of M4, no natal kicks were applied to black holes, but the
population had decreased from about 1000 to one by an age of about
9Gyr (see \citet[Fig.15]{HG2008}), and the last black hole escaped
before 12Gyr.  The last black holes were expelled from our model of
NGC6397 even earlier.  The contrast with our model of 47 Tuc is
striking.  Though natal kicks (with a 1-dimensional dispersion of
190km/s) were applied, and the number of retained black holes
decreased abruptly to 34 (near the left margin of Fig.\ref{fig:nbh}), more than half were still present at 12Gyr. 

\red{The three examples in Fig.\ref{fig:nbh} already raise an
  interesting question:  how is it that our model of NGC6397, which
initially  retains far more stellar-mass black holes than 47 Tuc, ends up
at the present day with none, while our model of 47 Tuc still has an appreciable
population?  The answer is that it depends on the phase of dynamical
evolution in which the cluster is found.  Our model of NGC6397
undergoes a phase of core collapse which ends at about the time when
the last stellar-mass black hole escapes \citep{GH2009}, but for our
model of 47 Tuc this phase of core collapse lies far in the future
\citep{GH2011}.  We return to this issue in Section
\ref{sec:bh-and-evolution}, remarking here only that we refer to this
episode as  {\sl second core
  collapse}, to distinguish it from a very early phase in which the
mass segregation of the 
system of black holes comes to an end.}

\red{With reference to Fig.\ref{fig:nbh},} our aim in this paper is to provide similar theoretically-based
expectations for the globular cluster M22 (NGC6656), motivated by the
recent discovery in that cluster of two stellar-mass black holes
\citep{St2012}, which may even represent only a sample of a
considerably larger population.  From what has been said, our first
step will be to construct a Monte Carlo evolutionary model which, like
those of the other three clusters we have studied, resembles the star
cluster at the present day.  This we do in the following section by
iterating on the initial conditions so that, after 12 Gyr of
evolution, the model provides an approximate fit to the observed
surface brightness and velocity dispersion profiles of the cluster,
and to its local stellar mass function (or, strictly, luminosity
function).  We repeat the exercise for different assumptions about the
natal kicks of stellar-mass black holes, in each case reporting the
number which survive to 12 Gyr (Sec.\ref{sec:BHpop}).  Our final section summarises our
conclusions, and discusses them in the context of other recent
research.  

\section{A Monte Carlo model of M22}

\subsection{Observational data}\label{sec:observations}

Our source for the surface brightness profile of M22 is the compilation of
\citet{Tr1995}.  As will be seen in Fig.\ref{fig:sbp}, the data are quite
scattered (by about half a magnitude) within the core, and
(surprisingly) a little fainter inside the core than at the edge of the core.

For the velocity dispersion profile we have adopted results from
\citet[their Fig.3]{La2009}.  The cluster rotates, with a maximum projected
  rotation velocity of about 3km/s \citep[\red{their} Fig.2]{La2009}.  It is not
  clear whether this has been removed from their velocity dispersion
  profile.  In any event the rotation is one dynamical property of the
  cluster which cannot be modelled with the existing Monte Carlo
  technique; this is confined to non-rotating and spherically
  symmetric systems.

\begin{center}
  \begin{figure}
  \includegraphics[width=9cm]{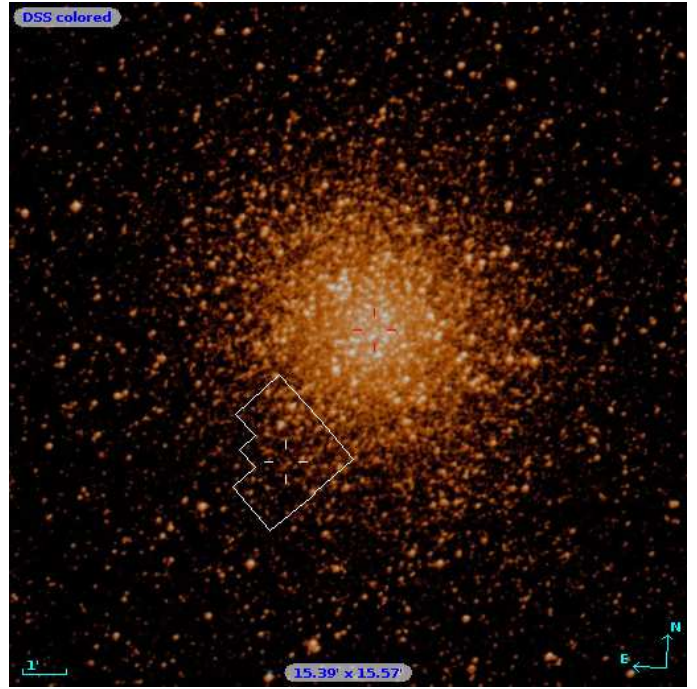}
  \caption{\red{The field where the luminosity function of
      \citet{PZ1999} was obtained, overlaid on an optical image of M22.  The
    luminosity function was obtained from only part of the WFPC2 field illustrated.}}
  \label{fig:hst}
\end{figure}
\end{center}

The stellar luminosity function we have used is the $V$ luminosity
function at a projected radius of about 4.5 arcmin given by
\citet{PZ1999}. 
This paper gives the data for each magnitude bin as
the number of stars per `HST area'; since the luminosity functions
have been obtained from a single WFPC2 chip \red{(see Fig.\ref{fig:hst})}, we take this to be an
area of 1.78 arcmin$^2$ {\red{(based on data in  the online {\sl HST Data Handbook for
  WFPC2})}}.   \red{As the core radius is about 1.33 arcmin
  \citep{Ha1996}, it is possible that appreciable mass segregation is
  present within the observed field, but only one luminosity function
  is given.  In principle, the Monte Carlo model should be mapped to
  the observed field, but we have simply compared the observed
  luminosity function with that in the
  Monte Carlo model at the radius of the centre of the field.}

Most other data on M22 (distance, metallicity, extinction) have been
taken from the on-line revision of December 2010 of the Harris
catalogue \citep{Ha1996}.  The only exception is the binary abundance,
where we have adopted a value of around 0.05 from the detailed results
of \citet{Mi2012}.  Those results cover the region within the
half-mass radius, and the\red{re is no statistically significant
  difference in the}  fraction  inside the
core.  In fact changing the binary fraction within reasonable limits
makes no \red{appreciable} difference to the overall evolution.

\subsection{Model assumptions}\label{sec:model}

The initial conditions we have adopted are similar (except of course
for several numerical values, given in Secs.\ref{sec:ics} \red{and
  \ref{sec:full-size}}) to those used in our previous papers
(see especially \citet[Table 1]{GH2011}).  Briefly these are King
models, with no
initial mass segregation.  The single stars have a two-part power law initial mass
function in the range from 0.1 to 100$M_\odot$, while the initial properties of the binaries are
taken from \citet{Kr1995}.   The Galactic tide is implemented as
described in \citet{GHHH2013}, but in any implementation it has to be treated as static in the
Monte Carlo model.  Note that \citet{Di1999} give for M22 an orbit with apo-
and peri-galactic distances of about $R_a = 9.3$ and $R_p =2.9$kpc, respectively,
corresponding to an eccentricity (defined as $e =
(R_a-R_p)/(R_a+R_p)$) of $e \simeq 0.52$.
There is, however, substantial evidence from $N$-body simulations that
a cluster on such an orbit retains a remarkably steady profile
throughout each orbit \citep{Ku2010}, and loses mass at a rate like a
cluster on a circular orbit at a fixed intermediate radius \citep{BM2003}.
  For the age of the cluster we have
adopted 12Gyr, though it could be even older \citep{MF2009}; and
for the metallicity we have taken $Z = 0.0004$ \citep{Ha1996}.
\red{Newly born neutron stars are given a kick using a Gaussian
  distribution with a one-dimensional dispersion of 190km/s, except
  for one model (Model C) reported in Sec.\ref{sec:full-size} and
  subsequent sections of the paper,  for which the dispersion was
  253km/s.  Natal kicks for black holes are more uncertain, and
  different choices are discussed (in connection with three models,
  called A,B,C), in Sec.\ref{sec:BH}.}

\subsection{Initial parameter values}\label{sec:ics}

The procedure now is to choose parameters specifying the initial
conditions (see Table \ref{tab:models}) so as to optimise the fit of
an evolved Monte Carlo model (Sec.\ref{sec:model}) to the
observational data (Sec.\ref{sec:observations}).  In our previous papers,
this has been a prolonged and laborious search, a process of trial and
error guided by intuition.  Since then it has been substantially
automated.  Our procedure now is to use the results of scaled models,
i.e. Monte Carlo models with a number of stars $N_\ast$ much smaller
than the number of stars in the star cluster, but adjusted so that the
relaxation time of the scaled model is the same as that of the actual
cluster (see \citet[Sec.2.4]{HG2008}).   Then a measure of goodness of
fit is constructed along $\chi^2$ lines; for each kind of data
(surface brightness, velocity dispersion and luminosity function) we
adopt a single measure of the dispersion of the error, one for each
kind of data, and based on information in the sources quoted in
Sec.\ref{sec:observations}.    These are normalised by the number of
data points in each kind of data, and simply summed, giving a measure
$Z$ of goodness of fit.  {(The Monte Carlo data are also
  subject to sampling error, but this has been ignored in the
  construction of $Z$.)}   Given values of
the seven adjustable parameters in Table \ref{tab:models} (i.e. $N,r_t,r_h,W_0,\alpha_1,\alpha_2,m_b$), we run the Monte Carlo
code and compute $Z$.  To optimise over the parameter space we employ
the Downhill Simplex algorithm, coded as {\sl amoeba} in
\citet{PTVF1992}.  For purposes of brevity in this paper we refer to this procedure
as {\sl ICFind}.

It is remarkable that the method is successful, as the algorithm is designed for
optimisation of a smooth function, whereas the results of the Monte
Carlo code are stochastic.  Nevertheless it appears to converge, from
a wide variety of starting points, after computation of order 50-100
models.  For $N_\ast = 10^5$ this takes a few days.  By
``convergence'' here we mean that the code finds a best model which
cannot be improved on in the number of iterations stated; from other
starting points, the best model may well be different.  Unfortunately,
even with this automatic method, we do not have any quantitative way
of deciding the range of acceptable models, which would require
improvement of our procedure for defining and calculating $Z$, and
much greater computational effort.
Furthermore, while computing the last 50 or so models, our experience is
that the code evolves models with very similar initial conditions and
chooses the best; it is, in effect, sampling the distribution of
models which all result from these initial conditions by different
choices of random numbers.  

\begin{center}
  \begin{figure}
  \includegraphics[width=9cm]{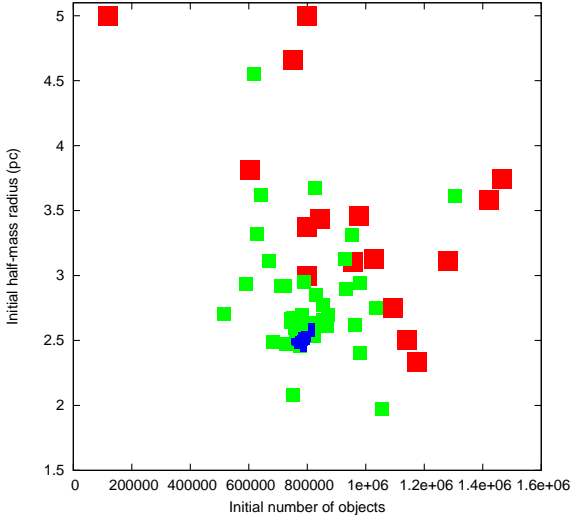}
  \caption{Convergence of the initial values of $N$ and $r_h$, where
    $N$ is the number of objects (single stars and binary stars) and
    $r_h$ is the half-mass radius, for the determination of initial
    conditions in the case of 100\% retention of black holes.  The plot gives
    an impression of the range of values sampled by the code.  \red{Large
    symbols give the first 20 iterates, medium-sized symbols give
    iterates 21--60, and small symbols give the remainder (101
    iterations altogether).}}
  \label{fig:convergence}
\end{figure}
\end{center}

\begin{center}
  \begin{figure}
  \includegraphics[width=9cm]{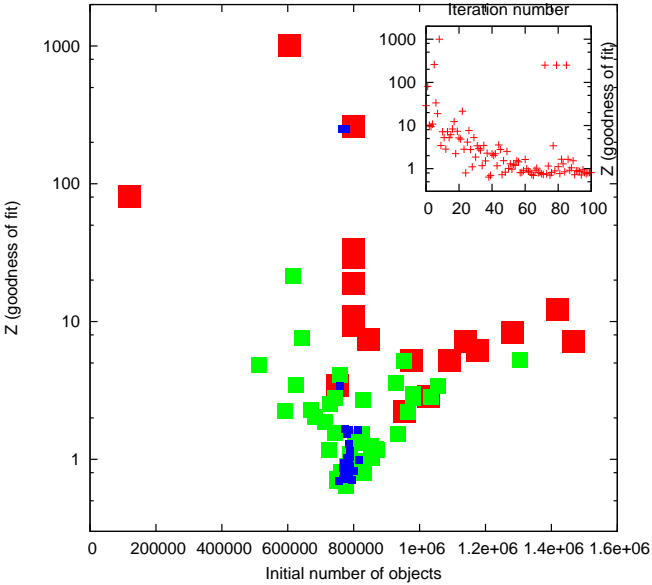}
  \caption{Marginal dependence of $Z$ (a measure of goodness of fit)
    on the initial value of $N$ (the initial number of objects), for the determination of initial
    conditions in the case of 100\% retention of black holes.  Default large
    values of $Z$ may occur if the model failed to reach the required
    age of 12Gyr.  Large values may also occur close to the
    best-fitting value of $N$ (about $7.8\times10^5$) if other
    parameters are far from optimal.  \red{The meaning of the symbols is
    given in the caption to Fig.\ref{fig:convergence}.  The inset
    gives the evolution of the goodness-of-fit parameter with
    iteration number.}}
  \label{fig:convergence-of-Z}
\end{figure}
\end{center}

Results for 100\% expulsion of black holes by natal kicks, and for 100\%
retention, are given in columns 2 and 3 of
Table \ref{tab:ics}, referred to as Models a and c, respectively.  From what has been said, it is not known whether
the differences between these initial conditions are significant.
Nevertheless it can be argued that the cluster will expand less with
100\% expulsion \citep{MPPZH2004,MWDG2008}, and therefore the larger value for the initial
half-mass radius $r_h$ in this case is what would be expected.
Figs. \ref{fig:convergence} and \ref{fig:convergence-of-Z} give an
impression of the convergence of the run with 100\% retention.  We
defer to the following subsection a discussion of how well the models
agree with the observational data, as the full-sized models discussed
in the next section are the focus of subsequent examination of the
black hole population and its evolution.  

\red{The status of the listed retention factors requires some comment.  
We do not directly control this number, i.e. by ensuring that a
given fraction of black holes are expelled.  In our simulations, on
the other hand, it is
an indirect outcome of other choices, such as the dispersion of
kicks (see Sec.\ref{sec:BH}), as well as of the evolution of the model (which affects the escape
velocity, for instance).}  %

\begin{table}
\caption{Initial conditions for M22, and the resulting black hole population}
  \begin{tabular}{ccc|ccc}
Model&a&c&A&B&C\\
\hline
$N_\ast/10^5$&$1$&$1$&$9.21$&$8.32$&$7.57$\\
$N/10^5$&$8.0$&$7.8$&$8$&$8.32$&$7.57$\\
$r_t$ (pc)&93&100&77&89&102\\
$r_h$ (pc)&3.1&2.5&2.67&2.72&2.43\\    
$W_0$&3.7&2.9&6.0&7.4&2.93\\
$\alpha_1$&1.1&0.90&1.12&1.21&0.90\\
$\alpha_2$&2.7&2.7&2.43&2.72&2.8\\
$m_b$&0.73&0.67&0.84&0.96&0.67\\
\hline
$N_{BH0}$&--&--&1799&675&\red{450}\\
Retention factor&0\%&100\%&0.1\%&62\%&100\%\\
$N_{BH12}$&--&--&2&14&\red{43}\\
$N_{SBH12}$&--&--&1&7&\red{39}\\
$N_{BHBH12}$&--&--&0&2&\red{1}\\
$N_{BHNS12}$&--&--&0&0&\red{0}\\
$N_{BHWD12}$&--&--&0&1&\red{0}\\
$N_{BHMS12}$&--&--&1&2&\red{2}\\
\hline
  \end{tabular}
\label{tab:ics}

Explanation:
\begin{enumerate}
  \item $N_\ast =$ the actual number of objects (single stars plus
    binary stars) in the model
  \item $N =$ the number of objects
    when the model is scaled to M22
  \item $r_t =$ initial tidal radius in parsecs
  \item $r_h =$ initial half-mass radius in parsecs
\item $W_0 =$ initial value of the scaled central potential of a King model
\item $\alpha_1,\alpha_2,m_b$: parameters of the initial mass
  function, which is a two-part power law with powers $m^{-\alpha}$,
  where $\alpha = \alpha_1$ for mass $m < m_b$, and $\alpha =
  \alpha_2$ above $m_b$.
\item $N_{BH0} =$ number of stellar-mass black holes formed in the
  normal course of stellar evolution.
\item Retention fraction = fraction of black holes remaining after the
  escape of those escaping as a result of natal kicks.
\item $N_{BH12} =$ number of stellar-mass black holes remaining at
  12Gyr.
\item $N_{SBH12},N_{BHBH12},N_{BHNS12},N_{BHWD12},N_{BHMS12} = $
  number of single black holes, black hole-black hole binaries, black
  hole-neutron star binaries, black hole-white dwarf binaries, and
  black hole-main sequence star binaries, respectively, at 12Gyr.
\item[] { Note: some other data on model B at 12Gyr are given in Sec.\ref{sec:de}.}
\end{enumerate}\label{tab:models} 
\end{table}

\subsection{Full-sized models}\label{sec:full-size}

The initial conditions in \red{the second and third columns of} Table \ref{tab:ics} were determined with
small-scale models, but results like these from {\sl ICFind} have also been used as
the basis of a number of full-scale
models, i.e. models in which $N=N_\ast$.  In some of these full-sized
models the values given by {\sl ICFind} were adjusted manually when it was judged that this
might improve the fit with observations.  For example the slope of the
lower mass function might be altered when it was judged that this
would further improve the fit with
the luminosity function.  Apart from $N_\ast$, the other main
difference between the full-sized models and those from {\sl ICFind} is that the runs have %
been carried out with a more advanced version of the Monte Carlo code, which is
called MOCCA and is described in \citet{GHHH2013}.  The essential
differences for the present purpose are (i) that escape of particles
is modelled more closely on our current understanding of escape in
tidal fields (rather than through the use of a tidal cut-off), and (ii)
that three- and four-body interactions are calculated with a few-body
code {\citep[{\sl Fewbody}, see][]{Fr2004}} instead of with cross sections.  Since the mass-dependence of the
cross sections used in the older code is based on theory, and the
masses of black holes represent an extreme situation, it might be expected
(because of the second of these changes) that the results could differ
significantly from those of the older code.

So far we have computed almost 40 full-sized models, each of which takes a
few days, though a few failed for technical reasons before reaching 12
Gyr (our assumed age for M22).  The best of these, as judged by
comparison with the observational data,   use the initial
conditions in the last three columns of Table \ref{tab:ics},
i.e. Models A, B and C.  The basis of the initial
conditions was a set of earlier runs of {\sl ICFind} than those
which produced Models a and c.  
Note that the values of $N_\ast$ (the initial number of objects in the
model) and $N$ (the assumed initial number of objects in M22) are
equal in Models B and C; in these full-scale runs we have generally not optimised over the choice
of $N$, though Model A is an exception.  For Model B, the quality of the fit to the observational data is displayed in
Figs.\ref{fig:sbp} -- \ref{fig:lf} and is discussed in detail in the
following paragraphs, along with abbreviated comments about Models A
and C.

\subsubsection{Surface brightness profile}\label{sec:sbp}

\begin{center}
  \begin{figure}
    \includegraphics[width=9cm]{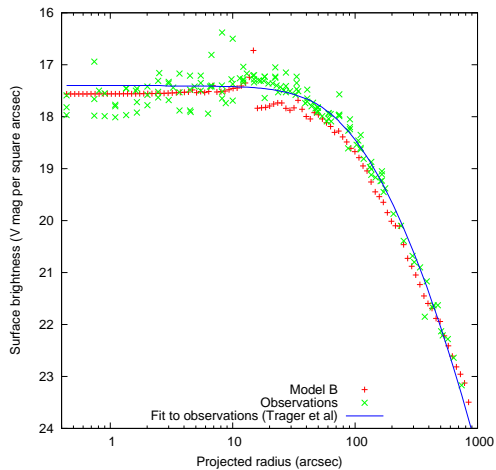}
    \caption{The surface brightness distribution of the Monte Carlo
      model (made with the new version of the code called
      MOCCA), with initial conditions in Column 5 of Table
      \ref{tab:ics}, compared with data from \citet{Tr1995}.  The
      solid line is their Chebyshev fit to the observation data.
      \red{The result for the Monte Carlo model is obtained by
        treating each star as a spherical shell, and projecting the model on the
        sky.  When a line of sight passes just inside the shell of a
        bright star the inferred surface brightness shows
     a spike, as in this model at a projected radius of about 15 arcsec.}   }
\label{fig:sbp}
  \end{figure}
\end{center}

 The surface brightness profile of the model is compared in
Fig.\ref{fig:sbp} with the observational data from \citet{Tr1995} and
a Chebyshev polynomial fit which they provide.  The model is generally
somewhat fainter than the observations, by about 0.3 mag.  It looks
particularly faint at the edge of the
core, especially when compared with the actual observational data and
not to the smooth fit to the observational data.  The  mismatch looks
considerably smaller in the halo, but  the profile there is much
steeper.

For Model A the surface brightness  (not shown) is close to 17.5 up to
a projected radius of 20 arcsec, too faint (by up to 0.5 mag in
places) from there  up to about
100 arcsec, and quite satisfactory thereafter.  \red{The surface
  brightness of Model C matches the observational data rather well
  (within the size of the symbols in Fig.\ref{fig:sbp}).}



\red{While this discussion has been expressed in terms of a comparison of
surface brightness at a given radius, other interpretations are
possible.  For example, a model which exhibits an underluminous halo
may simply be one that is too small (in radius).}


\subsubsection{The projected velocity dispersion profile}

\begin{center}
  \begin{figure}
    \includegraphics[width=9cm]{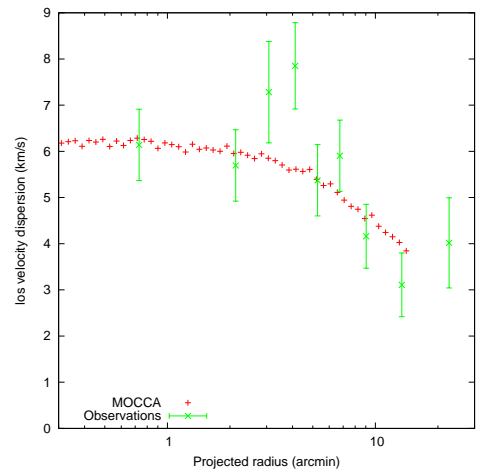}
    \caption{The line-of-sight velocity dispersion profile of the same
       model as in Fig.\ref{fig:sbp}, compared with data from
       \citet[\red{their} Fig.3]{La2009}.  The model data are plotted only up to
       the outermost radius of the observational surface brightness
       profile shown in Fig.\ref{fig:sbp}.   }
\label{fig:vdp}
  \end{figure}
\end{center}

 It
  is hard to characterise the fit of the velocity dispersion profile
  of Model B (Fig.\ref{fig:vdp}) with confidence,
  because of the large scatter in the observational data, but it may
  be best summarised by saying that it would be hard to improve.  
Perhaps the subjective impression is that the  velocity dispersion of
the  model is a little too small, but a number of factors should be
borne in mind.  First, the outermost point includes stars close to the tidal
radius (about 32 arcmin, according to \citet{Ha1996}), where velocity
dispersion profiles are elevated by the effects of the tidal field
\citep{Ku2010}\footnote{ This is especially true around
  perigalacticon.  Note that the current Galactocentric distance of
  M22 is 4.9kpc \citep{Ha1996}, i.e. about one third of the way from
  peri- to apogalacticon.}, and
these effects are not included in the Monte Carlo models.  Second,
even with a binary fraction of 5\%, the velocity dispersion may be
elevated by the internal motion of binaries.  Third, membership was
determined on the basis of two spectral line indices, the radial velocity,
and projected distance from the cluster centre, which led to the
inclusion of only 345 stars out of the total of 3407 spectra, and so
interlopers may still exist.  Fourth, the typical uncertainty in the
radial velocity of an individual star is about 3kms$^{-1}$.  On the
modelling side, it is also important to recall 
that the Monte Carlo model ignores the rotation of the cluster.

The projected velocity dispersion profile for Model A is very similar
to that for Model B (just described), but for Model C the result is
\red{noticeably poorer.  Though the central line-of-sight velocity
  dispersion is satisfactory at about
7 kms$^{-1}$,  at larger radii it falls more steeply than the result
for Model B shown in Fig.\ref{fig:vdp},  closely following the lower
envelope of the observational data outside about 5 arcmin.}


\subsubsection{The local luminosity function}\label{sec:lf}

\begin{center}
  \begin{figure}
    \includegraphics[width=10cm]{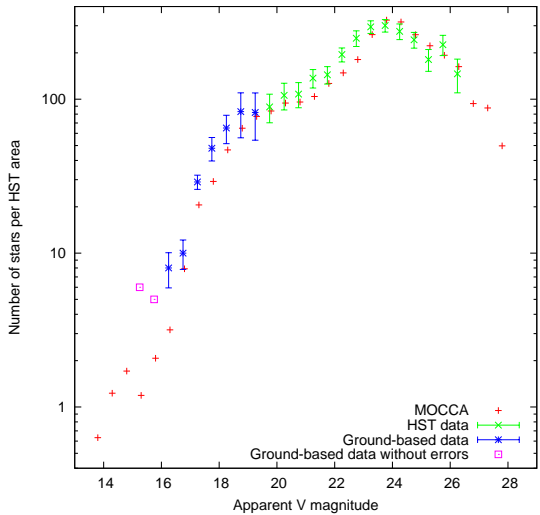}
    \caption{The luminosity function  of the same
       model as in Fig.\ref{fig:sbp}, compared with data from
       \citet{PZ1999}.  Though some of this is ground-based, all the
       data is scaled to their HST field.  The error bars, where available, have been
       read from their Figs.9 and 11.}    
\label{fig:lf}
  \end{figure}
\end{center}

Compared with the observational data, the luminosity function of Model
B (Fig.\ref{fig:lf}) shows a deficit of stars brighter than turn-off
($m_V\simeq 18.4$) and in a section of the main sequence.  At the
brightest magnitudes the deficit may reach 0.5 dex or more, though the
absence of error estimates in the observational data here makes this
uncertain.  In the rest of the main sequence the agreement seems
satisfactory, especially in the absence of any estimate of the
uncertainty in the Monte Carlo prediction.  These results may go some
way to explaining the fact that the surface brightness of the model is
generally a bit too low (Sec.\ref{sec:sbp}).


Since the agreement near turn-off seems satisfactory, 
it is
difficult to  understand how the deficit can rise so much  within the
small mass-range of stars brighter than turn-off, unless there is some flaw in
the stellar evolution package in post-main sequence evolution.
Similarly, it is difficult to know how the agreement along the main
sequence could be improved by varying the mass function index
$\alpha_1$ (see Table \ref{tab:ics}).  

The mismatch of the luminosity function of Model A to the
observational data is of a similar magnitude to that for Model B, but
is qualitatively different.   The model has an excess of stars
brighter than
about magnitude 25, and a deficit at fainter magnitudes.  Model C
\red{is qualitatively similar to Model B, except that the fit to the
  observational data is slightly worse around magnitude $m_V \simeq
  18$ and around $m_V = 25$, but fits better between these limits
  .}

\subsubsection{Dynamical evolution}\label{sec:de}

\begin{center}
  \begin{figure}
    \includegraphics[width=10cm]{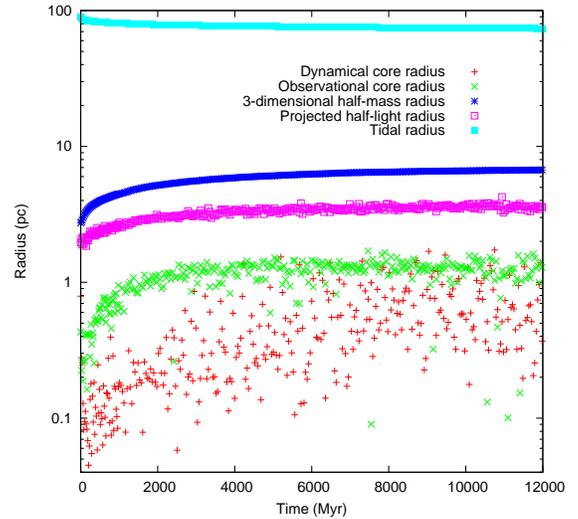}
    \caption{The evolution of the tidal radius, two versions of the
      half-mass radius, and two versions of the core radius, for Model
      B.}    
\label{fig:radii}
  \end{figure}
\end{center}

{ As we shall see in Sec.\ref{sec:CD}, the dynamical evolutionary phase
of a cluster is one of the main factors in assessing its likely
population of stellar-mass black holes, and so we discuss the
dynamical evolution of Model B here.  

The initial mass of the model is about $5.70\times10^5M_\odot$, and
shrinks to about $3.20\times10^5M_\odot$ at 12 Gyr.  The resulting
modest decrease in the tidal radius is shown in Fig.\ref{fig:radii}.
The value at the present day, about 73.6pc, greatly exceeds the
observational value \citep{Ha1996} of about 29.7pc, but these can mean
very different things in the case of a model which underfills its
tidal radius, as here:  the initial edge  radius of the King
model is 25.0pc, compared with the initial tidal radius of 89pc (Table
\ref{tab:ics}).
The value of the tidal radius of  Model B is consistent with the
Galactic potential.  As mentioned in Sec.\ref{sec:model}, the Galactic orbit of the cluster  takes
it between 3 and 9 kpc from the Galactic Centre \citep{Di1999}, and 
in a
Galactic potential with a flat rotation curve at 220km/s the range of
tidal radii is from about 50 to 105pc, which includes the value in the
model, but not the observational one.

Fig.\ref{fig:radii} also shows two versions each of the core and
half-mass radii, i.e. a theorist's version and an observer's one.  The
observer's values compare well with those given in \citet{Ha1996},
which are about 1.24pc and 3.12pc for the core and half-light radii,
respectively.  
}

\section{The population of stellar-mass black holes}\label{sec:BHpop}

\subsection{Evolution of total numbers}\label{sec:BH}

Finally we turn to the main motivation of our study, which is the
number of stellar-mass black holes at 12Gyr.  Depending on the slope of the
upper mass function, the number of stellar-mass black holes formed is between
about 200 and 900 as the slope of the upper mass function, $\alpha_2$, is
decreased from 3.0 to 2.6.  The numbers for the three best full-sized
models are given in Table \ref{tab:ics} and are labelled as $N_{BH0}$,
although of course it is not the initial number.  Though the numbers
vary widely, this is almost entirely explained by the variation in $\alpha_2$.

\begin{center}
  \begin{figure}
    \includegraphics[width=10cm]{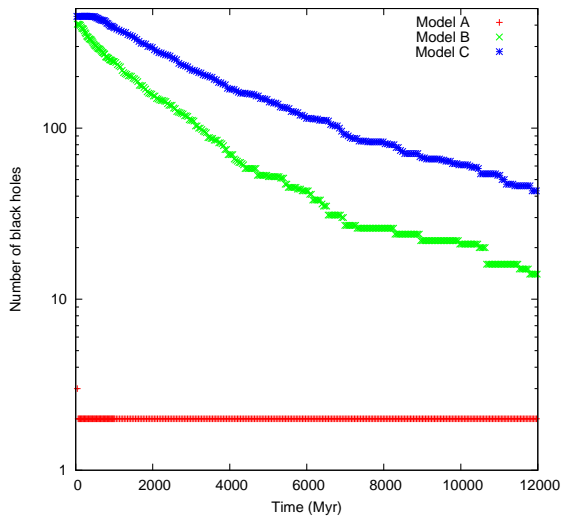}
    \caption{The evolution of the number of stellar-mass black holes
      in three models.  For two models the initial number formed lies
      outside the plotted range, but is given in Table \ref{tab:ics}.
}    
\label{fig:nbhABC}
  \end{figure}
\end{center}

{ The subsequent evolution of the number of black holes depends
crucially on the primordial kicks given to all new black holes, as
these three models illustrate.  
If all new black holes are given a kick with a
1-dimensional dispersion of 190kms$^{-1}$ (as is commonly considered
for neutron stars), almost all escape promptly, and  very few are still present at 12 Gyr.  This is illustrated in Fig.\ref{fig:nbhABC} by the
data for Model A, and the final number is listed as $N_{BH12}$ in
Table \ref{tab:ics}.  For this particular quantity the result has not
been scaled from $N_\ast$ to $N$, since the scaling factor is nearly unity.
}

{   In the absence
of natal kicks, exemplified by Model C in Table \ref{tab:ics} and
Fig.\ref{fig:nbhABC}, the fraction of all stellar-mass black holes surviving at 12
Gyr is about \red{10}\%.   Similar values were obtained in  the
small-scale models produced by {\sl ICFind}.

The other recipe for natal kicks of black holes that we tried is the fall-back
procedure of \citet{BKB2002}, which applies no kick if a large amount
of mass from the supernova envelope falls back onto the degenerate
remnant.  Model B is the best of the full-sized models in which this
procedure was adopted, and 
the fraction of
black holes remaining at 12Gyr was about 2\%.
}

\subsection{The black hole population at 12Gyr}

\begin{center}
  \begin{figure}
    \includegraphics[width=10cm]{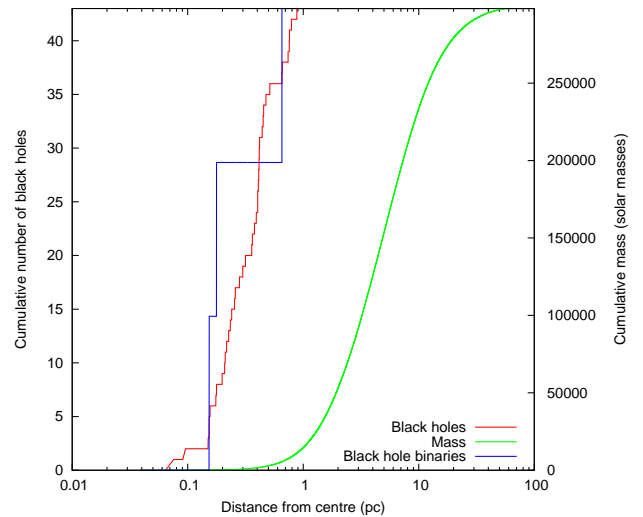}
    \caption{The spatial distribution of the black holes and black
      hole binaries \red{in Model C} at 12Gyr, compared with the spatial distribution
      of all stellar mass (including black holes).  No axis labels
      or ticks are given for the number of black holes binaries (i.e. binaries
      in which at least one component is a black hole); the curve
      represents the cumulative fraction of the number, scaled to the
      vertical size of the frame.  \red{For comparison, the two black holes
    discovered by \citet{St2012} lie at {\sl projected} radii of 0.25 and
    0.4pc, respectively.}}    
\label{fig:bhdist}
  \end{figure}
\end{center}

{ As may be expected from the effects of mass segregation, the spatial
distribution of the black holes at 12Gyr is very centrally
concentrated (Fig.\ref{fig:bhdist}, \red{which shows data for Model
  C}).  \red{The spatial distribution of the few}
binaries in which at least one component is a black hole \red{is
  statistically indistinguishable}.  In
addition to two-body interactions leading to mass segregation, such
binaries are subject to energetic dynamical interactions which can
send the binary into the halo of the cluster, but there is no evidence
from Fig.\ref{fig:bhdist} that this is noticeable in their spatial
distribution. In Model B  the outermost black hole binary is at
0.35pc from the centre.  The projected distances from the centre of M22
of the two black holes found by \citet{St2012} are 0.25pc and 0.4pc.
}

Table \ref{tab:ics} gives more detail on the numbers of black hole
binaries, broken down according to the type of the companion.  Again
as expected (from the effect of exchange interactions and their
mass-dependence) the companions \red{tend to be} drawn from the relatively
high-mass populations in the model.  Because of the interest in
finding a source for the emission of observed black holes in M22,
details are given in { Table 2} 
of the companions and
orbital parameters.  The binary identifier gives the model (Table
\ref{tab:ics}) and the order in increasing radial distance within the
model.  

Unfortunately not one of these binaries is close to Roche-lobe
overflow.  { Though the data correspond to conditions at 
a time close to 12Gyr, we have also checked that none of the black hole binaries
 are accreting at
any time in the period 10--12Gyr, in any of the three full-sized
models A--C.  \blue{From the total of 59 black holes in these models, it follows that the probability that a black hole
  is accreting from a binary companion is at most a few percent.
  Therefore, if it is assumed that the two black holes in M22 have
  Roche-lobe filling companions, model A can be ruled out, i.e. the
  model in which all black holes experience natal kicks similar to
  those of neutron stars.  This conclusion is model-dependent,
however, for reasons given in the next paragraph and in Sec.\ref{sec:limitations}.}}

Despite the evidence of models A--C, we also checked three
other models with rather similar initial parameters, and found
altogether four examples of black holes accreting from evolving
stellar companions.  According to the models, accretion continued for
at least 0.5Gyr, and so the probability \blue{that a black hole is
  accreting from a binary companion}  
\blue{could indeed be a few percent}.  It is, however,
difficult to estimate this probability, especially as it is likely to
depend on the choice of parameters for the primordial binary
population, and we have not attempted to explore this.





\begin{table*}

\begin{minipage}{166mm}
\begin{center}
\caption{Black hole binaries in models of  M22}
\end{center}
  \begin{tabular}{rccccccccc}
Binary&A1 &B1&B2&B3&B4&B5&C1&C2&C3\\
Primary mass ($M_\odot$)&10.0&12.6&10.0&11.0&10.0&8.8&\red{14.6}&\red{9.0}&\red{3.5}\\
Companion mass ($M_\odot$)&0.23&0.78&9.9&0.69&10.0&0.64&\red{10.0}&\red{0.33}&\red{0.61}\\
Companion type&MS&MS&BH&MS&BH&WD&\red{BH}&\red{MS}&\red{MS}\\
Companion radius ($R_\odot$)&0.24&1.02&--&0.70&--&0.012&\red{--}&\red{0.31}&\red{0.57}\\
Semi-major axis ($R_\odot$)&312&177&106&132&399&449&\red{112}&\red{157}&\red{161}\\
Eccentricity&0.52&0.62&0.28&0.44&0.99&0.38&\red{0.84}&\red{0.46}&\red{0.80}\\

  \end{tabular}
\end{minipage}

\end{table*}

\section{Conclusions and discussion}\label{sec:CD}

\subsection{Conclusions}

\red{Motivated by the recent discovery of two stellar-mass black holes
  in the Galactic globular cluster M22, we have constructed dynamic
  evolutionary models of this object in order to assess the survival
  of its population of black holes to the present day.  We find that
the
result depends heavily on the assumptions made about natal kicks
applied to new stellar-mass black holes.  For kicks with a one-dimensional
dispersion of 190km/s, the number of stellar-mass black holes at the
present day is no more than one or two (Model A in Table
\ref{tab:ics}).  
If no kicks are applied, then
the fraction remaining at the present day is of order 0.1, resulting
in a number of order 40 (Model C).  Model B represents an
intermediate, but physically motivated assumption about natal kicks,
and results in a present-day population numbering 14.}

\red{
We computed the dynamical evolution of our models with a  Monte Carlo method.  This code, of which we used two
versions, includes two-body relaxation, binaries and their dynamical
interactions, escape in the Galactic tide, and procedures for the
internal evolution of both single and binary stars.  Using a new
procedure, we have explored hundreds of sets of initial conditions so
as to produce models which, after 12Gyr of simulated evolution,
resemble M22 in their surface brightness profile, velocity
dispersion profile and stellar luminosity function.  Possible
initial conditions obtained by this procedure are summarised in Table
\ref{tab:ics}, and Figs.\ref{fig:sbp}--\ref{fig:lf} compare one of the evolved models  with the
observational data.}


\subsection{Black holes and cluster evolution}\label{sec:bh-and-evolution}

 It is useful to try to draw some general lessons about the surviving
black hole populations in old globular clusters from the modelling of
M22 described in this paper, and from similar models of a few other
objects, summarised here in Fig.\ref{fig:nbh}.  Some, like M4 and
NGC6397, lose all, or almost all, of their black holes well before the
present day, while others (47 Tuc and M22) retain a\red{n appreciable}
fraction (assuming\red{, in the case of M22,} that natal kicks are moderated in some way).  These
facts are related to the evolution of the core.  As we have seen
(Fig.\ref{fig:radii}) the core of M22 shows no sign of collapsing
yet.  Even the very concentrated cluster 47 Tuc is no more than
half-way to core collapse \citep{GH2011}.  Of the four clusters which
are under discussion, these are also the two with \red{appreciable} residual
populations of black holes (provided that kicks do not eject almost
all new black holes).

The link between black hole populations and the evolution of the core
has been noticed before \citep{MPPZH2004}, and is underpinned by
a recent theoretical treatment by \citet{BH2013}.  These results show that
expansion of the core (and indeed of the half-mass radius) can be driven by dynamical interactions among
the black holes, which inevitably lead to their escape.  Eventually
the population of black holes is insufficient to sustain the flow of
energy by relaxation in the outer parts of the cluster, and then the
core begins to \red{contract}.  \red{As a result of this, energy is increasingly
  generated by interactions between the remaining black holes and the
  stars of lower mass, and the rate of escape of black holes
  declines.  This change can be seen in Fig.\ref{fig:nbhABC} at about
  the time when the core radius reaches its largest values
  (Fig.\ref{fig:radii}).  }
This phase of core \red{contraction} ends at what \citet{BH2013} call
``second core collapse'' (the first being the original collapse of the
black hole subsystem), when some other mechanism of generating energy
(e.g. primordial binaries) becomes efficient enough.  

\red{The evolution of our model of M22 is more complicated than that of the
idealised models considered by \citet{BH2013}, but does not differ
qualitatively.  Indeed, though}
  stellar
evolution also contributes to the early expansion of the half-mass
radius,  \citet{GH2011} showed that primordial binaries make little
difference at the early stages \red{(in their Monte Carlo model of 47 Tuc)}.  

The upshot of th{\red{ese discussions} is that \red{appreciable} populations of
stellar-mass black holes are only to be expected in clusters which
have not yet passed (second) core collapse.  Other things being equal,
this means clusters which have a sufficiently long evolutionary time
scale, and we note that the half-mass relaxation times of NGC6397 and
M4 are under 1Gyr ($\log t_{rh}=8.60,8.93$, respectively, according to
\citet{Ha1996}), while those of M22 and 47 Tuc exceed 1Gyr ($\log
t_{rh}=9.23,9.55$, respectively).

These considerations allow us to synthesise not only our modelling of the four
globular clusters that we have discussed, but also two other recent
studies.  
\begin{enumerate}
\item \citet{MUFR2013} also used a Monte Carlo code to study the
problem, though the model was not specifically geared to M22.  The
retention factor was high, about 86\%, and more than half of the retained
black holes still survived in the cluster at 12Gyr.  We estimate the
half-mass relaxation time at 12Gyr to be about $6\times10^9$yr, though
this is based on the half-mass radius, whereas the estimates above are
based on the half-light radii.  Estimating these radii from Model B (Fig.\ref{fig:radii}), we
find that the comparable value of the half-mass relaxation time is
about $2\times10^9$yr, a little larger than the value for M22.
\item  The
other model which we mention here is a direct $N$-body model
\citep{SH2013} with $N = 2.5\times10^5$ initially, and a similar
binary fraction to our Monte Carlo models.  Though smaller than M22 in
mass, the larger initial radius of the $N$-body model gives it a
value for the relaxation time at 12Gyr of about 2.1Gyr.  The initial retention
fraction was 10\%, but even so 16 remained at 12Gyr.  Naively, this
would scale to about 50 for an initial model comparable in size to our
suggested initial conditions for M22.  
\end{enumerate}
}

{ Despite this tidy picture, mention must be made of M62, which has
  a recently announced black hole candidate \citep{Ch2013}, despite an
  uncomfortably low relaxation time: $\log t_{rh} = 8.98$.}

\subsection{Limitations of the modelling}\label{sec:limitations}

Now we consider some aspects of the models which could have a bearing
on these conclusions.  In the first place we can make no claim for the
uniqueness of the initial conditions we have derived.  In particular,
if more compact initial conditions exist (i.e. with a smaller
half-mass radius), then the central escape velocity would be higher
than in the existing models (for example, about 57km/s at the start of
Model B), and
the retention fraction of black holes would be greater, under any reasonable
hypothesis on the magnitude \red{of natal kicks}.

More problematic are aspects of the evolutionary history of globular
clusters which are not modelled at present in the Monte Carlo code.
It has recently been suggested \citep{LBMP2013} that accretion of
interstellar gas (slow ejecta from stellar evolution) will act as a
resistive force on the motions of black holes.  This takes us to perhaps the
most popular scenario for the formation of second generations in
Galactic globular clusters (see, for example, \citealt{DE2008}), in
which the ejected gas sinks to the centre of the original cluster of
first-generation stars, and forms a second generation, while much of
the first generation escapes.  No evolutionary simulation yet includes
these complex processes, and one can only argue qualitatively about
how this may affect our conclusions.  

In this scenario it is often
argued that the first generation may be more massive than the second.
Therefore the cluster in which the first generation of stellar-mass black holes
formed would have had a much higher escape velocity than we have
envisaged, making retention of a large fraction of these black holes much more
likely.  They would be centrally concentrated (by mass
segregation), and would not be expected to escape, unlike much of the
rest of the 
first generation.  Equally, it is hard to see how the survival of black holes
in the second generation would be adversely affected by being immersed
in the potential well of the remaining first generation.
\blue{Finally, these considerations suggest that sufficient numbers of black
  holes might well survive to the present day in this scenario, even if they were
  subject to natal kicks as in our model A.}

\section*{Acknowledgements}

We thank Jay Strader for guidance on the choice of observational data
on M22, \red{and the referee for his comments, which have markedly improved our efforts}.
This work was partly supported by the Polish Ministry of Science and
Higher Education through the grant N N203 38036, \red{and by the National Science Centre through the grant
DEC-2012/07/B/ST9/04412}.

\bsp

\label{lastpage}

\end{document}